# Berry phase shift from 2π to π in Bilayer graphene by Li-intercalation and sequential desorption

Ryota Akiyama[*,1], Yuma Takano[1], Yukihiro Endo[1], Satoru Ichinokura[1], Ryosuke Nakanishi[1], Kentaro Nomura[2], and Shuji Hasegawa[1]

[1]*Department of Physics, The University of Tokyo, 7-3-1 Hongo, Bunkyo-ku, Toyko 113-0033, Japan*

[2] *Institute for Materials Research, Tohoku University, 2-1-1 Katahira, Aoba-ku, Sendai, Miyagi 980-8577, Japan*

Abstract

We have found that the Berry phase of bilayer graphene becomes π from 2π estimated by Shubnikov-de Haas oscillations when the A-B stacked pristine bilayer graphene experiences the Li-intercalation and sequential Li-desorption process in ultrahigh vacuum. Furthermore, the mobility of such processed bilayer graphene increases around four times larger, ~ 8,000 cm$^2$/V·s, than that of the pristine bilayer graphene. This is mainly due to increment of the scattering time and decrement of the cyclotron mass, which can be interpreted as a result of the change of the stacking structure of bilayer graphene from A-B to A-A, corresponding to a change from the parabolic to the linear band dispersion.

---

[*] akiyama@surface.phys.s.u-tokyo.ac.jp

Monolayer graphene is well known to be a two-dimensional (2D) Dirac system [1-3]. However, bilayer graphene (BLG), which is composed of two layers of graphene stacking together, can be both Dirac and Schrödinger systems, depending on its stacking structure. When BLG is in the A-B stacked form, in which the upper graphene is shifted in-plane by $(\vec{a}+\vec{b})/3$ with respect to the lower one, it forms a 2D Schrödinger system having a parabolic band dispersion [4] (where $\vec{a}$ and $\vec{b}$ are the unit vectors of graphene sheet). On the other hand, when it is in the A-A stacked form, in which the lattice of two layers is stacked without lateral shift, it forms a 2D Dirac system having a linear band dispersion [5-7]. This difference between the A-B and A-A stacked BLGs, of course, causes some difference in their physical and electronic properties. Especially, the Berry phase is different between the two systems; that of Dirac system is π, while that of Schrödinger system is 2π.

In spite of these remarkable characteristics of BLG, the method of fabricating the A-A stacked BLG has not been reported yet except for the local observation [8,9]. This is because the A-B stacking is energetically more favorable than the A-A stacking [10]. In a recent study, however, Li-intercalated BLG was found to take the A-A stacked form with Li atoms intercalated at the center of the hexagons in the upper and lower graphene sheets [11]. Bulky graphite Li-intercalation compound (GIC) is also known to be in the A-A stacking [12,13].

In this study, it is suggested, based on *ex situ* measurements of Shubnikov-de Haas oscillations, that the A-A stacked BLG grown on a SiC substrate can be obtained by desorbing Li atoms from Li-intercalated BLG with heating in ultrahigh vacuum (UHV) environment. The results offer a pragmatic suggestion of fabrication of the A-A stacked BLG.

The samples used in the present study were pristine BLG (Sample A), Li-intercalated BLG [14], and Li-desorbed BLG (Sample B). Their RHEED patterns are shown in Fig. 1 (a) − (c), respectively. Sample A was prepared on a *n*-type Si-rich 6H-SiC (0001) substrate by heating up to 1550°C in UHV ($3 \times 10^{-10}$ Torr). Because the heating temperature and the keeping time were optimized, we have successfully fabricated exclusively BLG. As seen in Fig. 1(a), Sample A showed both the graphene 1×1 pattern (indicated by red arrows) and the buffer layer 6√3×6√3 R30° pattern (indicated by blue arrows). For preparing Sample B, Li was first deposited on the BLG using a Li dispenser (SAES Getters) at room temperature in UHV. This Li-intercalated BLG sample showed a RHEED pattern of √3×√3 R30° (indicated by yellow arrows) as shown in Fig. 1(b) [14,10,15]. Such a RHEED pattern reflects Li atoms arranged regularly in the interlayer space. Here, the green arrows show the SiC substrate 1×1 pattern. Then, it was heated up at 900°C to desorb Li atoms until the √3×√3 R30° pattern disappeared and the 6√3×6√3 R30° pattern from the buffer layer revived as shown in Fig. 1(c). This is Sample B.

The samples were taken out from the UHV chamber, and bonded to Au wires with Indium to form the conventional 6-terminal methods. All electrical transport measurements were performed with physical property measurement system (PPMS, Quantum Design Inc.).

Figure 1(d) shows the temperature dependence of the sheet resistance $R_S$ of Samples A and B in the temperature range of 2 – 300 K. $R_S$ increases rapidly at ~ 100 K in both samples with decreasing temperature because carriers in the SiC substrate are frozen out. Thus, $R_S$ below ~ 70 K is not due to the substrate but solely to the BLG. The small increment of the $R_S$ at the low temperature < ~ 20 K is due to the weak (anti-)localization as reported in monolayer and BLG [16-18]. The top inset represents the magnetic field dependence of the sheet resistance at 2 K. To obtain the mobility by the semi-classical Kohler's law [19], the magnetic field dependence of resistance at $B > ~ 7$ T was fitted as the light green lines by $R(B) \sim 1 + (\mu B)^2$ where $\mu$ is the mobility and estimated to be 440 and 350 cm$^2$/V·s for Samples A and B, respectively. The bottom inset is the magnetic field dependence of the Hall voltage at 2 K. The estimated carrier density by the Hall effect of Sample A and B are $3.7 \times 10^{13}$ cm$^{-2}$ and $5.4 \times 10^{13}$ cm$^{-2}$ with $n$-type, respectively.

The two-dimensional material which has high mobility like graphene can be expected to show quantum oscillations such as a Shubnikov-de Haas (SdH) oscillation [20-25]. The results of SdH oscillation enable us to obtain the fundamental physical properties, such as the Berry phase, the cyclotron mass, the scattering time, and mobility. Since SdH oscillations appear in the longitudinal resistance $R_{xx}$, we plotted the inverse magnetic field dependence of $\Delta R_{xx}$ at some temperatures (2 – 20 K) in Samples A and B in Figs. 2(a) and (b), respectively. $\Delta R_{xx}$ was deduced by subtracting the background in $R_{xx}$ which was given by the third polynomial fit. The amplitude of oscillations decreases with increasing the temperature in both samples, and the oscillations in Sample A seem to decrease with temperature more rapidly than in Sample B. For analysis of the Berry phase, the Landau-level fan diagram is plotted in Fig. 2(c) for both samples. The peak (valley) in $\Delta R_{xx}$ was assigned to the Landau index $n+1/2$ ($n$). Then the intercept at $1/B = 0$ in this diagram can be estimated by fitting straight lines of $n = \alpha/B + \beta$ with the fitting parameters of $\alpha$ and $\beta$ by Lifshitz-Onsager quantization rule. When $|\beta|$ is close to 0.5 (0), the Berry phase is to be $\pi$ ($2\pi$). The intercept $\beta$ for Samples A and B was estimated to be $0.05 \pm 0.16$ and $0.55 \pm 0.04$, respectively. This result indicates that the Berry phase is shifted from $2\pi$ to $\pi$ by the Li-intercalation and sequential desorption process, resulting in the change of carriers from Schrödinger Fermions to Dirac Fermions. Figure 2(d) shows $dR_{xx}/dB$ vs. $1/B$ at 2 K, where the oscillations of Samples A and B are compared. The definite shift of the phase in the oscillation between two samples is seen, which directly corresponds to the shift in the Berry phase. Both oscillations were fitted to determine the periodicity by $dR/dB = a\sin(F/B+\phi)+b$ as indicated by dashed red and blue curves where the fitting parameters are $a$, $F$, $\phi$, and $b$. The resultant fitting values of $F$ in Samples A and B are 716 ± 3 T and 788 ± 6 T, and hence the carrier density was estimated to be $2.8 \times 10^{12}$ cm$^{-2}$ and $3.0 \times 10^{12}$ cm$^{-2}$, respectively.

Figures 3(a) and (b) show the oscillatory component of conductance $\Delta\sigma_{xx}$ at some temperatures (2 – 40 K) of Samples A and B, respectively. The subtracted background to deduce $\Delta\sigma_{xx}$ is the third polynomial fit for the longitudinal conductance $\sigma_{xx}$. The standard Lifshitz−Kosevich (LK) theory [26] gives the temperature dependence of the amplitude in SdH oscillation as

$$\Delta\sigma_{xx}(T)/\Delta\sigma_{xx}(0) = \lambda(T)/\sinh(\lambda(T)). \qquad (1)$$

Here, the thermal factor $\lambda(T) = 2\pi^2 m_{cyc} k_B T/\hbar eB$, where $m_{cyc}$ is the cyclotron mass and $k_B$ is the Boltzmann constant. The temperature dependence of the oscillation amplitude at $B = 10.4$ and 10.1 T in Samples A and B, respectively, are shown in Fig.

3(c), where the solid curves are the fitting results to Eq. (1). The estimated values of $m_{cyc}/m_e$ for Samples A and B are 0.13 ± 0.04 and 0.09 ± 0.01, respectively, where $m_e$ is the electron rest mass. The cyclotron mass of Sample B is ~ 30% smaller than that of Sample A.

With the above estimated $m_{cyc}$, the lifetime of carriers is given by the Dingle plot [27-29]. The Dingle theory describes the oscillations amplitude $\Delta R_{xx}$ as

$$\Delta R_{xx} = 4R_0 e^{-\lambda_D} \lambda/\sinh(\lambda), \qquad (2)$$

where $R_0$ represents the resistance at zero magnetic field and $\lambda_D = 2\pi^2 m_{cyc} k_B T_D/\hbar eB$. Here, $T_D = \hbar/2\pi k_B \tau$ is the Dingle temperature, where $\tau$ is the total scattering time. In Fig. 3(d), the inverse magnetic field dependence of $\ln[\Delta R_{xx}\sinh(\lambda)/4R_0\lambda]$ in Samples A and B at 2 K is shown. The solid lines indicate the linear fittings. The $\tau$ values were calculated from $T_D$ which was estimated by the slopes of the fitting lines in Fig. 3(d). The resultant values of $\tau$ are 0.11 ± 0.06 psec and 0.40 ± 0.20 psec in Samples A and B, respectively. The scattering time of Sample B is around four times larger than that of Sample A. Consequently, the mobility $\mu$ of carriers was deduced by the relation of $\mu = e\tau/m_{cyc}$ where $e$ is the elementary charge. The estimated values of $\mu$ are 1,900 cm$^2$/V·s and 8,000 cm$^2$/V·s in Sample A and B, respectively. All physical parameters estimated here are listed in Table 1.

These results can be attributed to a change in the band dispersion due to the change of stacking structure from the A-B to the A-A type. The band dispersion of the A-A type is linear, resulting in the Berry phase of $\pi$, a higher mobility and a lighter cyclotron mass. Since the Li-intercalated BLG is reported to be A-A stacked [11], our results suggest that the A-A stacking is dominant even after Li-desorption. Here, it should be noted that the semi-classical mobility estimated by the Kohler's law (~ 300 - 400 $cm^2$/V·s) is inconsistent with ones estimated by SdH oscillations in both samples. This is because the cyclotron radius is enough small (~ 20 nm) to detect only the local properties within domains whereas the semi-classical longitudinal resistance and the Hall effect reflect the properties over many domains in the whole transport path (~ 1 - 2 mm). Domain boundaries between different stackings and ripples in graphene sheets would be additionally induced by the Li-intercalation and desorption process. Therefore, the carrier density can be also affected by such domains. More specifically, since the SdH oscillation is more contributed by domains having higher carrier mobility and lower carrier density, the values of carrier density and mobility estimated by the Hall effect and the Kohler's law are different from those derived from SdH oscillations.

In summary, we observed SdH oscillations in a pristine BLG (Sample A) and Li-desorbed one after Li-intercalation (Sample B) on SiC substrates. Then, we found that the Berry phase of the A-B stacked BLG (Sample A) is changed from $2\pi$ to $\pi$ by the Li-intercalation and sequential desorption process (Sample B). This seems to be induced by the change in the stacking type from A-B to A-A. Accordingly, the mobility of Sample B becomes around four times larger than that of Sample A. This fabrication method of the A-A stacked BLG paves the way for intensive use of Dirac-Fermion BLG.

We are grateful to A. Takayama and R. Hobara for a fruitful discussion. This work was partially supported by a Grant-in-Aid for Scientific Research (A) (KAKENHI No. JP16H02108), a Grant-in-Aid for Young Scientists (B) (KAKENHI No. 26870086), Innovative Areas "Topological Materials Science" (KAKENHI No. JP16H00983, JP15H05854) and "Molecular Architectonics" (KAKENHI No. 25110010) from Japan Society for the Promotion of Science. R.A. and Y.T. contributed equally to this work.

**Figure Captions**

Fig. 1 (Color online)  The RHEED patterns of (a) pristine BLG (Sample A), (b) the Li-intercalated BLG, and (c) the Li-desorbed BLG (Sample B). The incident azimuth of electron beam is $[11\bar{2}]$ of the SiC(0001) substrate. The graphene 1×1 pattern (indicated by red arrows), the buffer layer 6√3×6√3 R30° pattern (blue arrows), the regularly-arranged-Li √3×√3 R30° pattern (yellow arrows), and the SiC substrate 1×1 pattern (green arrows) are shown. (d) Temperature dependence of the sheet resistance of Samples A and B. The insets: (top) The magnetic field dependence of the sheet resistance at 2 K. The light green lines are the fitting results by the Kohler's law. (bottom) The magnetic field dependence of the Hall voltage at 2 K.

Fig. 2 (Color online)  (a), (b) The inverse magnetic field dependence of the $\Delta R_{xx}$ in Sample A and B at some temperatures (2 – 20 K), respectively. (c) The Landau-level fan diagram of Samples A and B. (d) The plots of $dR_{xx}/dB$ vs. $1/B$ at 2 K in Sample A (red) and B (blue), respectively. The two dashed curves are fitting by sinusoidal functions.

Fig. 3 (Color online)  (a), (b) The magnetic field dependence of $\Delta\sigma_{xx}$ in Samples A and B at some temperatures (2 – 40 K), respectively. (c) The Temperature dependence of the oscillation amplitude in $\Delta\sigma_{xx}$ at 10.4 and 10.1 T for Sample A and B, respectively. The solid curves are the fit by the LK theory. (d) The relation between $\ln[\Delta R_{xx}\sin(\lambda)/4R_0\lambda]$ and $1/B$ at 2 K. The lines are the fitting with the theory of the Dingle plot in Samples A (red) and B (blue), respectively. From the slope of the lines, $T_D$ can be estimated.

TABLE I. The physical parameters estimated for Samples A and B at 2 K: the carrier density by SdH oscillations, the scattering time, mobility, the cyclotron mass, and the intercept $\beta$ in Landau-level fan diagram.

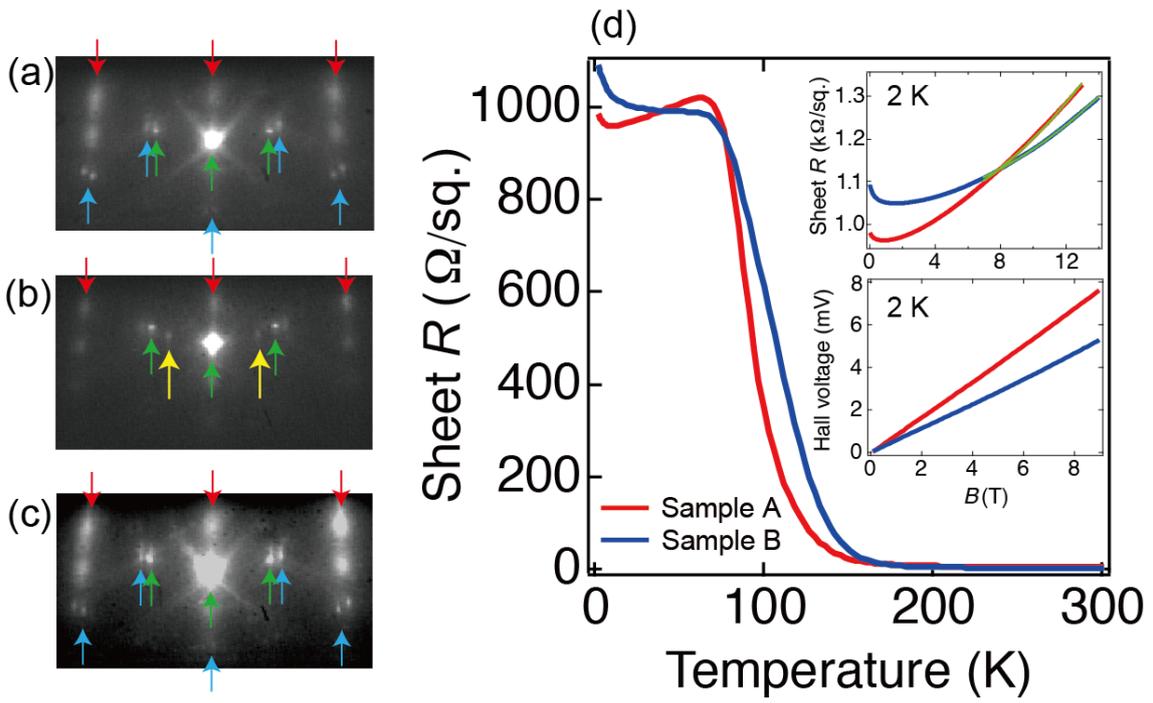

Fig. 1

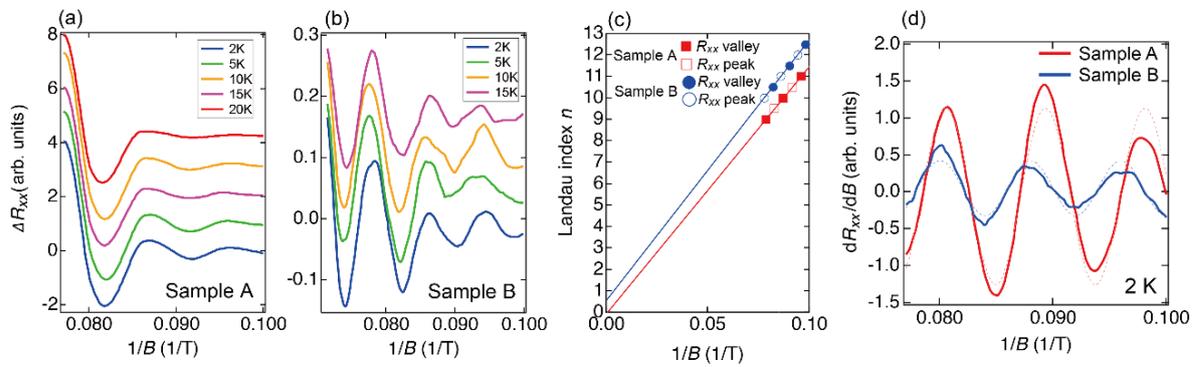

Fig. 2

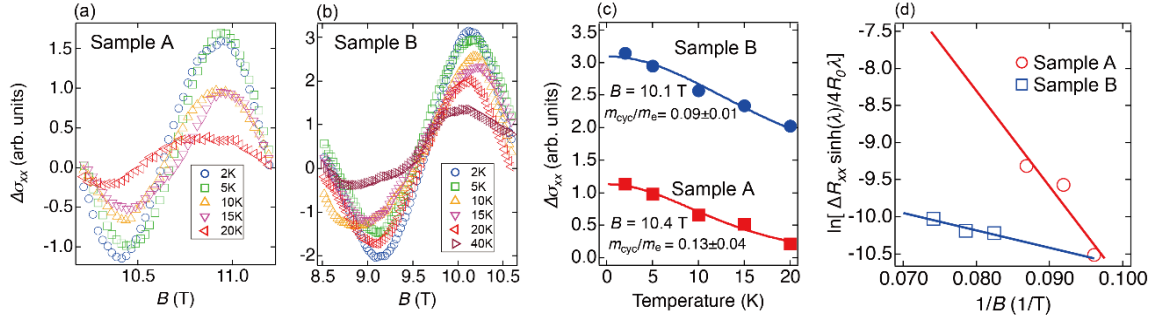

Fig. 3

TABLE I.

| Sample | A | B |
|---|---|---|
| Carrier density by the Hall effect (cm$^{-3}$) | $3.7 \times 10^{13}$ | $5.4 \times 10^{13}$ |
| Mobility by the Kohler's rule (cm$^2$/V·s) | 440 | 350 |
| Carrier density by SdH oscillations (cm$^{-2}$) | $2.8 \times 10^{12}$ | $3.0 \times 10^{12}$ |
| Scattering time (psec) | $0.11 \pm 0.06$ | $0.40 \pm 0.20$ |
| Mobility by SdH oscillations (cm$^2$/V·s) | 1,900 | 8,000 |
| Cyclotron mass ($m_e$) | $0.13 \pm 0.04$ | $0.09 \pm 0.01$ |
| Intercept $\beta$ in Landau-level fan diagram | $0.05 \pm 0.16$ | $0.55 \pm 0.04$ |